# Evaluating the capability of large language models to personalize science texts for diverse middle-school-age learners

Michael Vaccaro Jr.[1,*], Mikayla Friday[1], and Arash Zaghi[1]

[1]Department of Civil and Environmental Engineering, College of Engineering, University of Connecticut, Storrs, CT 06269

*Corresponding author*
*Email: michael.t.vaccaro@uconn.edu*

**Abstract**
Large language models (LLMs), including OpenAI's GPT-series, have made significant advancements in recent years. Known for their expertise across diverse subject areas and quick adaptability to user-provided prompts, LLMs hold unique potential as Personalized Learning (PL) tools. Despite this potential, their application in K-12 education remains largely unexplored. This paper presents one of the first randomized controlled trials ($n = 23$) to evaluate the effectiveness of GPT-4 in personalizing educational science texts for middle school students. In this study, GPT-4 was used to profile student learning preferences based on choices made during a training session. For the experimental group, GPT-4 was used to rewrite science texts to align with the student's predicted profile while, for students in the control group, texts were rewritten to contradict their learning preferences. The results of a Mann-Whitney $U$ test showed that students significantly preferred (at the .10 level) the rewritten texts when they were aligned with their profile ($p = .059$). These findings suggest that GPT-4 can effectively interpret and tailor educational content to diverse learner preferences, marking a significant advancement in PL technology. The limitations of this study and ethical considerations for using artificial intelligence in education are also discussed.

## 1 Introduction

In 2008, the National Academy of Engineering named advancements in Personalized Learning (PL) one of the fourteen grand challenges for the twenty-first century (National Academy of Engineering, 2008). Since this time, PL has emerged as a prominent area of education research. Through this work, PL has evolved into a broad term which now encompasses a vast number of interventions and programs (Shemshack and Spector, 2020; Walkington and Bernacki, 2020). The work presented in this paper aims to build on this existing research by investigating the potential of novel Large Language Models (LLMs) to foster highly adaptive PL environments.

Because of the existing variety in studies, numerous definitions of PL have been proposed. Each of these definitions differs in their incorporation of learner characteristics, instructional designs, and learning outcomes (Bernacki et al., 2021). This diversity can be problematic, as tools or interventions that meet the definition of PL in one framework cannot be guaranteed to do the same under another framework. Resultantly, efforts have been made to centralize the definition of PL. This study adopts the definition proposed by the U.S. Department of Education's Office of Educational Technology, which describes PL as "instruction in which the pace of learning and the instructional approach are optimized for the needs of each learner" (2017, p. 9). Specifically, learning activities should be self-initiated and relevant to students' interests, while learning objectives and the organization of educational content should be tailored to the individual (U.S. Department of Education, 2017).

Despite its promise (Pane et al., 2015), the widespread adoption of personalized learning in modern classrooms has remained limited. This is partly due to significant time and resource constraints, which pose logistical challenges that fundamentally prevent schools from individualizing materials for each student (Adelman and Taylor, 2018; Arnaiz Sánchez et al., 2019). In its simplest form, human-led PL would require each teacher to develop multiple versions of learning materials and assignments. Doing so would likely be a massive time-sink, especially given the wide variability in student needs and learning preferences (Sewell, 2022). These challenges are further exacerbated by the range of subject areas students are expected to master and schools' recent struggles to fill open positions (Bleiberg and Kraft, 2023). Together, they form crippling barriers to human-led PL in modern-day classrooms.

In an effort to overcome these barriers, researchers and educators have increasingly turned to educational technology. Although various technologies have been developed, such as online homework platforms and Intelligent Tutoring Systems (ITSs), they are typically specialized for particular topics or subject areas (Liu et al., 2017), limiting their applicability on a large scale. Recent advancements in artificial intelligence have opened a new frontier for PL. Specifically, modern LLMs such as ChatGPT have shown immense potential as educational tools. These models can quickly learn the unique needs and preferences of individual learners (Brown et al., 2020), allowing for the generation of personalized content in real-time. Furthered by the inter-disciplinary nature of their knowledge, LLMs hold an immense potential to push PL beyond the traditional bounds of recommender-style ITSs (Nunes et al., 2023; Chen et al., 2024b; Henkel et al., 2024). However, capable LLMs are just now coming onto the market and, as a result, their potential to personalize education has yet to be critically evaluated.

This study aims to fill this gap by conducting one of the first randomized controlled trials to investigate the role of LLMs in K-12 education. For this purpose, we evaluate OpenAI's fourth-generation Generative Pre-trained Transformer (GPT-4) model. Given the breadth of knowledge embodied in modern LLMs, we hypothesized that GPT-4 could foster a PL environment capable of adapting educational science texts to the learning preferences of individual middle school students. A small-scale study with 23 students from one Connecticut middle school was conducted to test this hypothesis. In this study, GPT-4 is used to adapt (i.e., rewrite) an educational text, in the form of a short selection from a textbook or article, to individual students' learning preferences identified along the dimensions of the Felder-Silverman model (1988). We evaluate the success of GPT-4 in this task using the Mann-Whitney $U$ test.

The remainder of this paper is organized as follows. The Related Work section introduces relevant research in PL and ITSs and discusses recent advancements in LLMs. Next, the Methods section describes the study design, participant pool, and the procedure used to collect data. The Results are presented next followed by a Discussion of their significance and the limitations of this study. Given that this study provides support for the integration of artificial intelligence (AI) and education, the Discussion section also notes several Ethical Considerations to be made in future works, such as the protection of student data. Finally, the Conclusion summarizes the contributions of this work and closes this paper.

## 2 Related Work

### 2.1 Intelligent Tutoring Systems

As implied by the Department of Education's (2017) definition, the goal of personalized learning is to adapt both the learning process and the educational materials to individual students' preferences and interests. PL interventions, especially those utilizing technology, have been investigated extensively for both K-12 and post-secondary education (Huang et al., 2016; Liu et al., 2017; Price et al., 2017; Hooshyar et al., 2021; Sibley et al., 2024). Many of these interventions fall under the umbrella of ITSs which, in general, refers to any system capable of tracking student achievement to support the learning process (Kulik and Fletcher, 2016). These systems have come in many forms, including both researcher-developed and commercially available software suites.

Intelligent Tutoring Systems have been shown to be effective learning tools, even when compared to human-led tutoring (VanLehn, 2011; Xu et al., 2019). For example, Contrino et al. (2024) used a commercial software to develop an adaptive learning environment for business students taking an introductory statistics course. Modest improvements in student grades were attributed to the adaptive technology. In another study, McCarthy et al. (2020) demonstrated how a Natural Language Processing model could be used to evaluate and improve high school students' reading comprehension. In their study, student reading comprehension was scored by an NLP model. These scores were then used to select the next passage shown to students. Researchers noted that this process provided students with meaningful scaffolds as the difficulty increased.

As Bulger (2016) points out, though, modern ITSs often act as "responsive systems" rather than as "adaptive systems." As such, ITSs are typically recommender-style systems that assign students to pre-determined materials in response to their performance rather than actively generating unique content aligned with their needs and learning preferences. Unlike traditional responsive ITSs, LLMs can be prompted with critical information about an individual learner. This information can come in several forms, ranging from examples of past performance to specific hobbies or interests (Chen et al., 2024a). In an attempt to move beyond the traditional decision-tree approach of responsive systems, this randomized controlled trial uses LLMs to develop a truly adaptive PL environment.

### 2.2 Large Language Models

OpenAI's GPT models have been applied in a wide range of fields, including psychology (Dhingra et al., 2023), neuroscience (Lee and Chung, 2024), and medicine (Waisberg et al., 2023), among others. Studies in these fields have largely been successful, which speaks to the diversity and wide range of knowledge captured in the models' training data (OpenAI, 2023). The multi-disciplinary success of these models can also be attributed to the transformer architecture which, in short, allows the LLM to parse the context of the input and generate highly-relevant responses (Vaswani et al., 2017; Bansal et al., 2024). While LLMs are known to make some mistakes with technical content, they have become more accurate over time (Katz et al., 2024; Roy et al., 2024). This trend is expected to continue as new models are developed and released.

Capitalizing on the power of these models, much research has focused on understanding the relationships between user-provided prompts and the generations that LLMs output. Much like how the success of a traditional machine learning model is dependent on model structure and selected features, the behavior of an LLM is determined by the nature and quality of the input developed by the model's user. Research of this kind has led to the development of a new field called prompt engineering. In this field, the focus is to develop a prompt, or a set thereof, that will lead to the best possible model output (Heston and Khun, 2023; Mizrahi, 2024). While some characteristics of what defines a good prompt are transferrable between use cases, prompts are often task specific. Because of this, prompt engineering has quickly evolved into a critical field of research. In fact, dedicated tools have recently been developed to assist researchers in crafting the most effective prompts for their specific use case (Arawjo et al., 2024).

Prompt engineering has been especially important in the application of LLMs to education, which is inherently multi-faceted. In education, prompts need to be designed in a way that accurate information can be generated across a range of use cases and subject areas. Recent research has demonstrated that LLMs can aid teachers by creating preliminary lesson plans (Hays et al., 2024), designing homework assignments or quizzes (Doughty et al., 2024), and automating the grading process (Kortemeyer, 2023; Latif and Zhai, 2024). In addition to serving teachers, research has aimed to explore the ways in which LLMs can be used to support students throughout the learning process, especially since these models are able to provide quick responses and appropriate scaffolds (Leinonen et al., 2023; Zhang and Tur, 2024). In a recent study, GPT has been found to be successful in adapting educational materials to match students' achievement level and evaluating students' performance on test questions in real-time (Jauhiainen and Guerra, 2023).

Despite the existing body of research dedicated to the possibilities for LLMs in education, there is a notable lack of experimental literature exploring the effects of these models on student learning. This is especially true for research in K-12 education, which has fallen behind research performed in post-secondary institutions. The present study, described below, aims to address this knowledge gap by being one of the first studies to experimentally investigate how well LLMs can personalize the learning process for individual middle school students.

## 3 Methods

### 3.1 *Study Design*

This randomized controlled trial aimed to investigate GPT-4's ability to personalize educational science texts for middle school students. In this study, the personalization was achieved by tuning GPT-4's outputs to align with students' individual learning preferences. This involved the careful design of the model's "system" and "user" prompts, which define its behavior and output (OpenAI, n.d.; Lee et al., 2024). These prompts were engineered by the research team through a previous study, discussed in Friday et al. (n.d.). In that study, simulations were run with several concurrent GPT-4 models to iteratively develop prompts that enable the model to systematically learn and generate text consistent with an individual's learning preferences. In these simulations, one model played the role of a human participant, one model described the learner, and a third model generated the personalized text. The latter two models were implemented in this study.

The present study is comprised of three components, as summarized in Figure 1. The first component, referred to as the training session, aims to identify the student's learning preferences. During this session, the student selects their preferred paragraph from a pair of science texts a total of four times. Each pair of paragraphs discusses the same topic and contains the same educational content. To capture a range of learning preferences in the training session, the paragraphs within each of the four pairs were rewritten using GPT-4 to target the extremes on one dimension of the Felder-Silverman learning style model. This model, updated in 2002, consists of four dimensions assessing how students best perceive (sensing or intuitive), receive (visually or verbally), process (actively or reflectively), and understand (sequentially or globally) information (Felder and Silverman, 1988). Because this study focuses on text, the visual/verbal dimension was replaced with an imagery/no imagery dimension. The paragraphs used in the training session were developed as part of the simulation study noted above (Friday et al., n.d.) and are provided on the author's GitHub page[1]. The topics of these pairs, chosen to be appropriate for middle-school, include the water cycle, climate change, photosynthesis, and the states of matter.

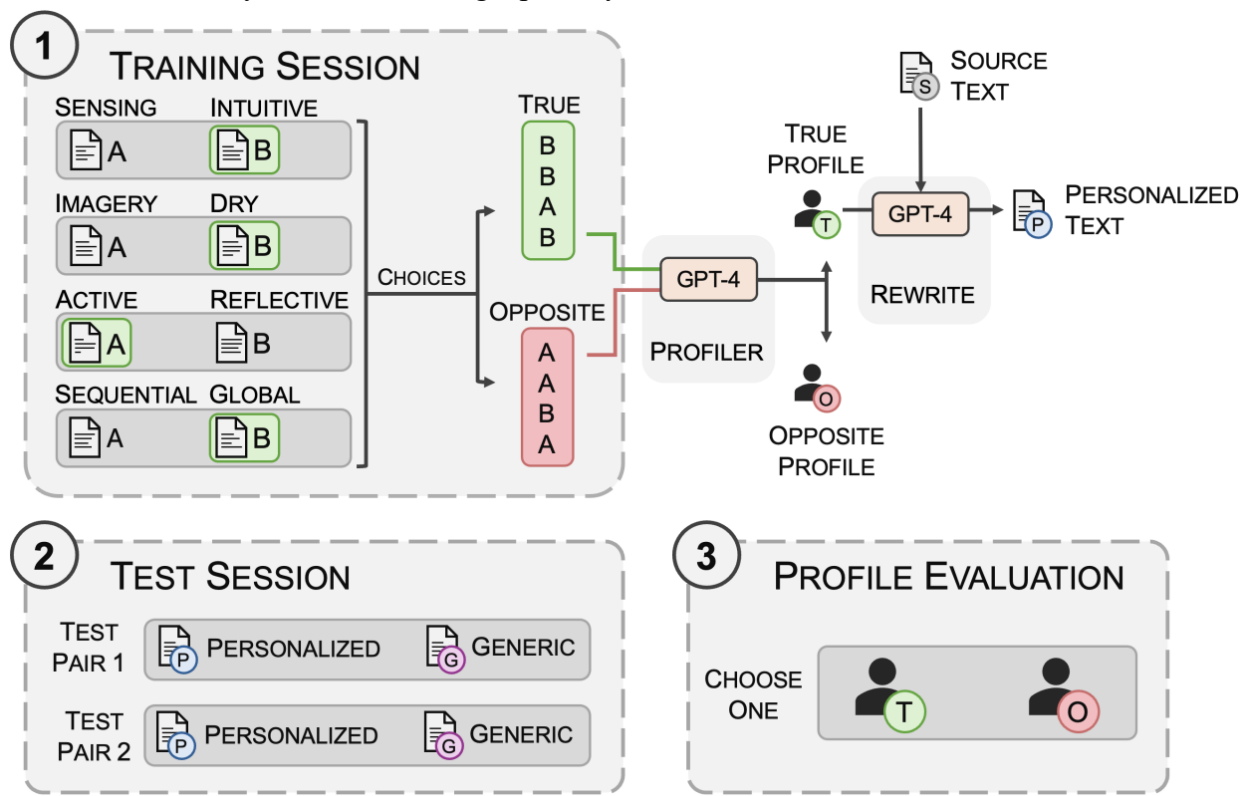

Figure 1. Three components of the study design used to generate and evaluate personalized texts with GPT-4, including the training session, test session, and profile evaluation. The process shown corresponds to the experimental group. Study components remain the same for the control group, except for the GPT-4 rewrite model, which takes the opposite profile as input in place of the true profile.

Next, data from the training session is fed into two GPT-4 models (*version: gpt-4-1106-preview*). The first model uses the choices made during the training session to generate a short description (or "profile") of the student's learning preferences, while the second model uses this profile to personalize an educational science-related text. The system and user messages that define these models, referred to as the "profiler" and "rewrite" GPT-4 models in Figure 1, were also developed in the study by Friday et al. (n.d.). The system and user messages for both models are available in the python code that defines them on GitHub.

---

[1] The author's GitHub repository for this study is at: https://github.com/m-vaccaro/LLMs-and-Personalized-Learning

Once generated, the personalized texts are tested against a generic text in the test session. To ensure consistency in content, the generic texts used in the test session were derived from the same source texts as those used to develop the personalized options. In addition, the generic texts were rewritten by GPT-4 to ensure consistency in the overarching tone of the two paragraphs in the test session. The generic texts were kept the same across participants as they were not adapted to any specific individual's learning preferences.

The final component of this study is the profile evaluation. In this phase, the accuracy of the student profile used to create the personalized texts is tested. Participants select between a true profile, generated according to their choices in the training session, and an opposite profile, generated against their choices in the training session.

Figure 1 outlines the study for the experimental group, where the personalized texts in the test session are created using the true profile based on the choices made in the training session. This structure is essentially unchanged for participants in the control group. In fact, the training session and profile evaluation phases are the same as those described for the experimental group. For the test session, though, the opposite profile is used to generate the personalized texts presented to the control group instead of the true profile. Thus, participants assigned to the experimental group were expected to select the personalized texts in the test session, while participants assigned to the control group were expected to select the generic texts.

## 3.2 Participants

A total of twenty-four seventh- and eighth-grade students (ages 11-13) from one middle school in Connecticut were recruited for this study. One participant withdrew from the study prior to data collection (final $n = 23$). An IRB-approved email and advertisement flyer was emailed to the parents and/or legal guardians of seventh and eighth grade students by school administrators to recruit participants. Hard copies of the flyer were also available for students to bring home. Emails and flyers contained a link to an online permission form which, once signed by a parent or legal guardian, enrolled students in the study. As this research involved minors, participants were required to provide assent at the time of their participation. Following their completion of all activities, participants were paid with either a pocket microscope or a scientific calculator. Participants were free to select the item they wished to receive, and both items valued near $20.

Participants were assigned to the experimental and control groups using stratified random sampling in which strata were formed on gender (Male/Female/Non-Binary). Twelve participants were randomly assigned to the experimental group and the remaining eleven participants were assigned to the control group. Although stratified random sampling was only conducted on gender, the experimental and control groups were well-balanced in terms of participant age, gender, and ethnicity. Table 1 summarizes these group demographics.

Table 1. Demographics of the experimental ($n = 12$) and control ($n = 11$) groups (Total $n = 23$).

|  | Grade | | Age | | | Gender[a] | | | Ethnicity[a] | | |
| --- | --- | --- | --- | --- | --- | --- | --- | --- | --- | --- | --- |
| Group | 7 | 8 | 11 | 12 | 13 | M | F | NB | White | Asian | Other[b] |
| Experimental | 5 | 7 | 0 | 4 | 8 | 7 | 4 | 1 | 8 | 3 | 1 |
| Control | 4 | 7 | 1 | 3 | 7 | 7 | 4 | 0 | 6 | 3 | 2 |
| **Total** | **9** | **14** | **1** | **7** | **15** | **14** | **8** | **1** | **14** | **6** | **3** |

[a]Gender and Ethnicity were reported by parents/legal guardians. M = Male; F = Female; NB = Non-Binary
[b]Other ethnicity includes "Asian/white" (Count: 1, Group: control), "Nepali" (Count: 1, Group: control), and "American Israeli" (Count: 1, Group: experimental)

## 3.3 Research Procedure

Data collection took place in the middle school to minimize barriers to participation like transportation. Students were scheduled into one 45-minute timeslot between 8 a.m. and noon by school administrators on one of five days. The study was conducted in an available office space and participants were not told which group (experimental/control) they had been assigned to. The office space featured a central round table where researchers and the participant could sit during the study. A separate desk was available for a school-approved faculty member who was present to supervise researchers' interactions with the students, as required by the IRB-approved study protocol and the school administration. Although present, the school faculty member was not involved in the study activities. Data collection took place between February and March 2024.

One student participated in the study at a time, and each student typically needed 25 to 35 minutes to complete the activities. After students assented to participation, computer-screen and audio recordings were started to capture any verbal responses or reactions from the participants and to serve as a back-up in case any issues occurred in saving. The study procedure contained in Figure 1 was implemented using a custom-built Graphical User Interface (GUI) developed in Python using CustomTkinter version 5.2.2 (Schimansky, n.d.), a publicly-available GUI library. This GUI ensured data were consistently formatted and saved between participants and avoided human-error in the data generation and collection processes. The GUI interacted with GPT-4 through OpenAI's Application Programming Interface (API) version 1.9.0 and implemented the study program for both groups. The full code for the GUI is available for download through the author's GitHub page.

A screenshot of the GUI during the training session is shown in Figure 2. The same layout was also used in the test session. The GUI was designed to have a simple and intuitive layout. In addition to stating the topic of the paragraph pair and providing the texts, a related image pre-generated with GPT-4 and was included to add some visual intrigue. The images used across the six science topics included in this study were the same for each participant. Finally, participants were provided with the simple instructions to "Please select your preferred paragraph from the two options below," with the gentle note to "Remember there is no correct response."

After the test session, participants were shown a screen which asked them to select the paragraph they believed provided the best description of their learning preferences. This corresponds to the profile evaluation phase of the study and is shown in Figure 3. Both the true and opposite profiles generated prior to the test session were shown, and participants were not told which were the true and opposite profiles. The location of the true profile, as either option 1 – left, or option 2 – right, was randomized for each participant. After making a selection, participants were asked to provide a brief justification. Because the session was also audio recorded, participants had the option of either writing out their response or stating their reasoning aloud.

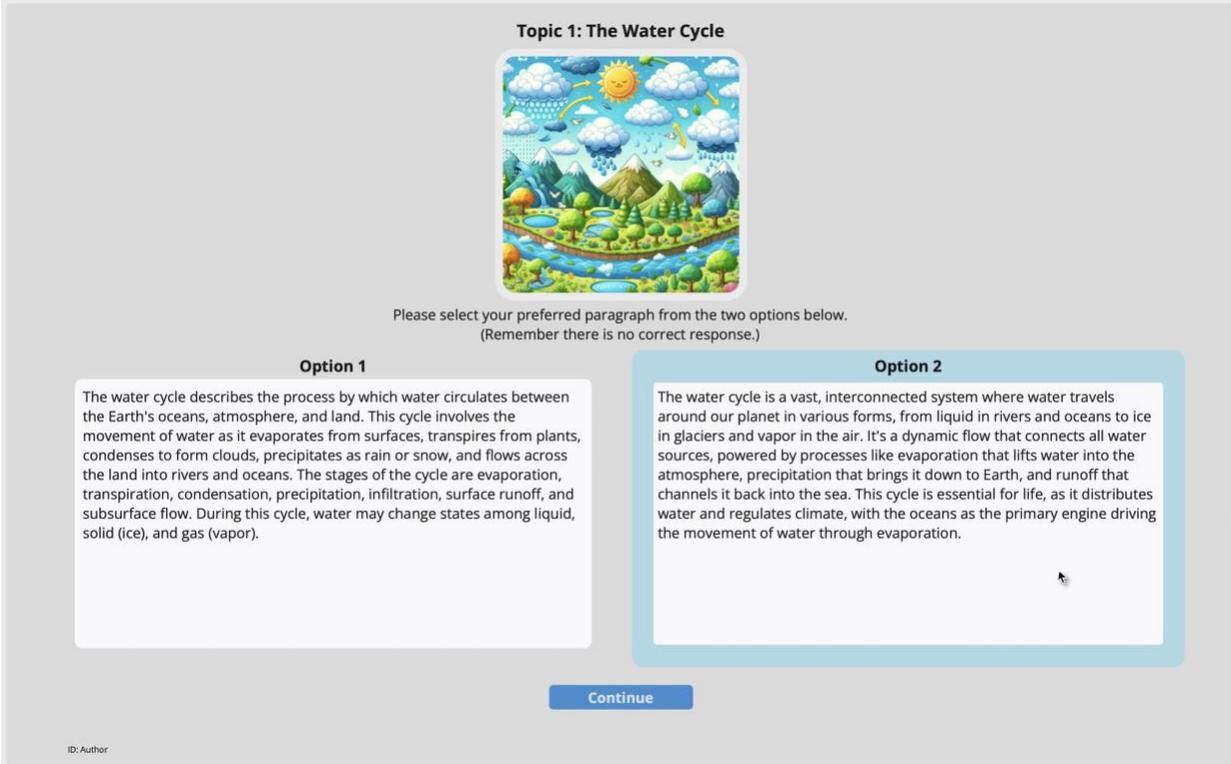

Figure 2. Screenshot of Graphical User Interface (GUI). One of the paragraph pairs used in the training session is shown (Topic: Water Cycle, Felder-Silverman Dimension: Sequential-Global).

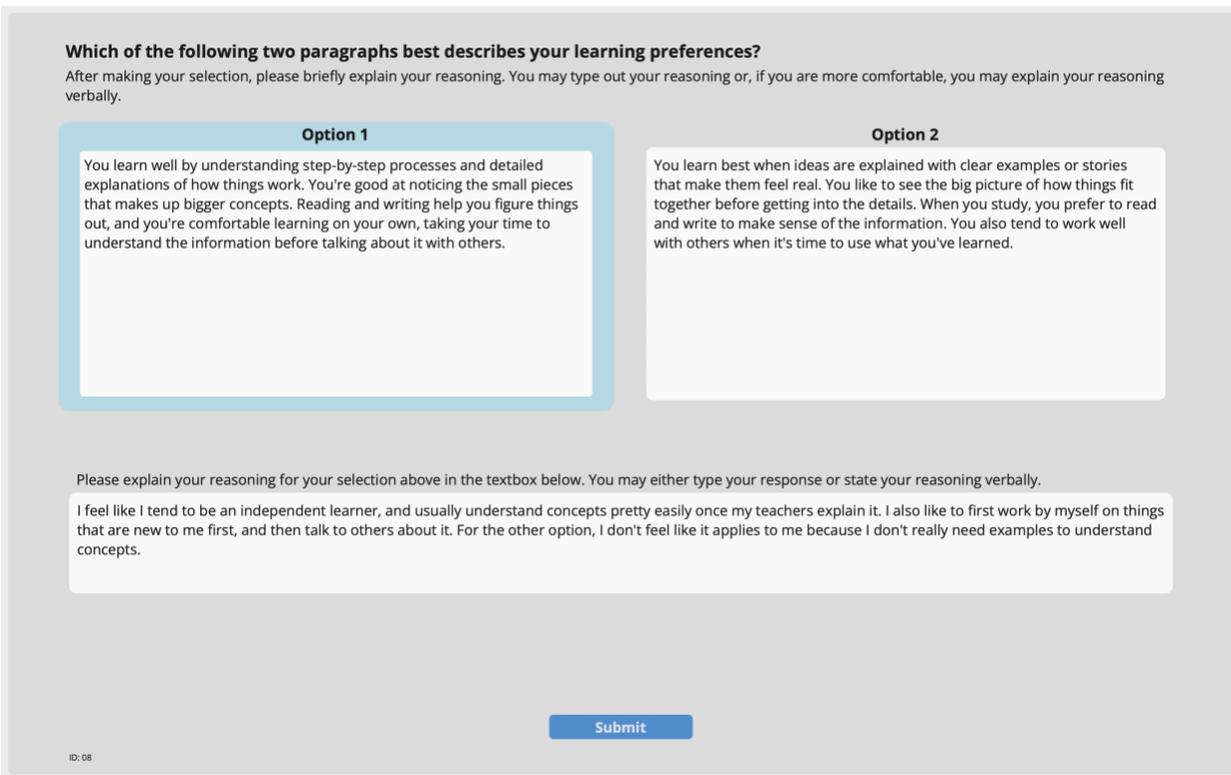

Figure 3. Profile Evaluation component of this study in the Graphical User Interface (GUI). Profiles shown are the true and opposite profiles generated by GPT-4 after the training session and before the test session.

## 4 Results

All analyses for this study were performed using IBM SPSS Statistics, version 29. All participants completed the study activities; thus, the dataset contains no missing or incomplete entries.

### 4.1 The Test Session – Effect of Personalization

A Mann-Whitney $U$ test was performed to assess the ability of GPT-4 to adapt text to the participant's learning preferences identified during the training session. A choice score was calculated for each participant based on data collected during their test session. A participant's choice score was equal to the number of times they selected the personalized paragraph over the generic text. Notice that participants in the experimental group were expected to select the personalized texts during the test session, while participants in the control group were expected to select the generic texts. If GPT-4 was successful, then, the average choice score of the experimental group should be high while that of the control group should be low. Because the test session consisted of two paragraph pairs, each participant's final choice score could be either 0, 1, or 2.

Table 2. Results of the test session comparing the number of personalized paragraphs selected by the experimental and control groups.

| Group | No. Selected Personalized Paragraphs | | | Total Score[a] | Mean | Std. Dev. |
|---|---|---|---|---|---|---|
| | 0 | 1 | 2 | | | |
| Experimental | 1 | 7 | 4 | 15 | 1.25 | 0.622 |
| Control | 5 | 4 | 2 | 8 | 0.73 | 0.786 |

[a]Total Score is the total number of personalized paragraphs selected during the test session (i.e., the sum of Choice Scores across all participants). The maximum possible total scores in the experimental and control groups are 24 and 22, respectively.

Table 2 summarizes the results of the test session for the experimental and control group. A Shapiro-Wilk test demonstrated that the data do not follow a normal distribution in both groups (Experimental group: $W(12) = 0.780$, $p = .006$; Control group: $W(11) = 0.799$, $p = .009$). Therefore, the Mann-Whitney $U$ test (Table 3) was used instead of the independent-samples $t$-test to compare the choice scores between the two groups (Wall Emerson, 2023). The mean choice score of the experimental group was found to be significantly greater than the control group at .10-level ($U = 40.5$, one-sided $p = .059$). These data indicate that participants in the experimental group were more likely to select the personalized paragraphs during the test session. If normality had been assumed and an independent samples $t$-test was used to compare the mean number of personalized paragraphs selected between the experimental and control groups, we find a medium-to-large effect size ($g = 0.715$) and a statistically-significant difference in means at the .05-level ($t(21) = 1.777*$, one-sided $p = .045$).

Table 3. Mann-Whitney $U$-test rank table.

| | Group | N | Mean Rank | Sum of Ranks |
|---|---|---|---|---|
| No. Selected Personalized Paragraphs | Experimental | 12 | 14.13 | 169.50 |
| | Control | 11 | 9.68 | 106.50 |
| | Total | 23 | | |

### 4.2 Profile Evaluation

Of the 23 participants enrolled in this study, 14 selected their true profile in the profile evaluation phase of this study (Table 4). This value is not significantly different ($p = .405$) from the expected value of the binomial distribution suggesting that, on average, participants performed no better than random when tasked with selecting their true profile. However, we do find a statistically significant difference at the .10-level in the proportion of participants who correctly selected their true profile between the experimental and control groups ($Z = 1.450$, one-sided $p = .074$, $h = 0.615$). Although greater than the traditional significance level of .05, this low of a $p$-value combined with a moderate effect size suggests that there may be some relationship between study group and the rate at which participants select their true profile in the final stage of the study. Since the profile evaluation phase occurred directly after the test session, it is possible that the personalized paragraphs implicitly influenced participants' choices. Specifically, it is possible that participants in the experimental group considered the features of the paragraphs shown in the test session when making their choice.

Table 4. Summary of data collected in the profile evaluation phase of the study.

|  | Selected Profile[a] | |
|---|---|---|
| Group | True Profile | Opposite Profile |
| Experimental ($n$ = 12) | 9 (0.75) | 3 (0.25) |
| Control ($n$ = 11) | 5 (0.455) | 6 (0.545) |

[a]Counts are shown. Values in parentheses indicate proportions.

## 5 Discussion

This study evaluated the ability of GPT-4 to systematically adapt educational STEM-related texts to the unique learning preferences of individual middle-school students. To evaluate this ability, participants in the experimental group selected their preferred paragraph between 1) a text adapted to their true profile and 2) a generic text, while those in the control group chose between 1) a text adapted to their opposite profile and 2) the same generic text. A Mann-Whitney $U$ test indicated that participants in the experimental group were significantly more likely to select the personalized paragraphs during the test session at the .10-level. These data suggest that GPT-4 can effectively adapt educational texts to the preferences of individual students and validates the system and user messages that define the two models. In addition, this result reinforces the notion that there is no "one-size-fits-all" text that is equally suited to the preferences of all learners, such as a single passage or article. This finding is consistent with the results of past research which has used LLMs to adapt text to students of varying academic achievement levels (Jauhiainen and Guerra, 2023).

To meet the definition of personalized learning, a learning environment must be capable of both pacing and tailoring the content of instruction to individual students and their personal interests (U.S. Department of Education, 2017). In this study, the Felder-Silverman learning style model was used to develop an initial profile describing each student's learning preferences. This profile was limited to the model's four dimensions, which do not explicitly account for a learners' preferred pace, specific interests, or existing knowledge. The profile also remained static during the experiment, limiting the amount of information that could be gleaned about each student to the four choices made during the training session. Despite these limitations, this study provides a valuable starting point for developing LLM-powered PL environments. Specifically, these profiles can be updated as students interact with the learning environment, refining their placement along the dimensions of the Felder-Silverman model and improving the LLMs overall performance.

It is important to recognize that we do not expect an LLM-powered PL environment to be implemented exactly as is presented in this randomized controlled trial. Beyond the creation of the initial profile, students' use of the learning environment would likely shift toward a more traditional chat-style interaction. Such an environment would allow the student profiles to move beyond the constraints of the Felder-Silverman model noted above. For example, students could ask the LLM as many questions as needed without the fear of judgement or the fear that the model will lose interest (Choi et al., 2024; Rogers et al., 2024). Furthermore, LLMs can quickly provide valuable feedback that can help keep students motivated, especially when they are engaged in difficult content (Meyer et al., 2024). Finally, further interactions between students and a LLM-based PL tutor could provide valuable insight into student learning (Chin, 2001). These insights could then be incorporated into the student profile. Such a dynamic interaction would likely lead to the development of more accurate profiles, steadily improving the quality of the text personalization over time.

Owing to their diverse capabilities and growing context awareness (Elyoseph et al., 2023; Bansal et al., 2024), LLMs also have the unique potential to support neurodiverse students (Addy et al., 2023). These students are often disadvantaged by traditional education systems, which do not accommodate diverse ways of thinking and learning (Chrysochoou et al., 2022). For instance, LLMs can help students with dyslexia by reorganizing or simplifying texts that may be difficult to read and comprehend, a strategy that has been shown to be beneficial (Alsobhi and Alyoubi, 2020; Rivero-Contreras et al., 2021). Research has also shown that PL may be extended beyond the adaptation of learning materials and into formative or summative assessments (Maya et al., 2021). In the context of this work, for example, student profiles may be used to inform the design of tests or homework assignments. This extension could offer students a more wholistic PL experience that situates all students on a truly equal playing field. This potential should be considered in future research as the role of LLMs in education continues to evolve.

## 5.1 Limitations & Future Research

While this study shows promise for the integration of LLMs in education, there are several key limitations that may affect the generalizability of the results presented above. First and foremost, this study is limited by its small sample size ($n = 23$). Despite emailing several middle schools and their administration, researchers only received approval to conduct research from one school. This greatly limited the population of students we could sample from. Although it would be preferable to collect more data, the use of this small sample in this study is justified by the restriction on the timing of data collection to be during the school year coupled with the rapid pace at which generative AI models are advancing (Bansal et al., 2024). As a result, our sample was limited to students who lived in a predominantly affluent region of the state. Some other factors leading to this small sample size may have included student self-selection and the level of comfort parents or legal guardians had with their child participating in an AI-based research study. In addition, our sample was limited to grade 7 and 8 students. Future research should aim to evaluate the broader efficacy of LLMs in personalizing education with larger samples and for a wider range of students before, during, and after the middle-school years.

The texts included in this research study were limited to middle school science topics and, more narrowly, to the water cycle, climate change, photosynthesis, the states of matter, plate tectonics, and electricity. These texts were intended to be introductory in nature so that they would be applicable to students of all achievement levels. As a result, students did not need to follow extensive lines of reasoning or solve problems mathematically. However, these are all abilities which should be incorporated into PL environments. Therefore, future research should further aim to ensure that LLM-based PL tools are broadly applicable in education and not specialized to one topic or one type of learning task.

Although this study provided support for the ability of GPT-4 to adapt texts to students' preferences based on their prior choices, the period of data collection was limited to a 25 to 35-minute session. In addition, this study did not assess differences in student learning or achievement between the experimental and control groups. More research is therefore needed to evaluate the long-term effects of LLM-powered PL environments on student success, as well as if the benefits of personalization observed in this study persist as students continue to use the platform over time.

### 5.2 Ethical Considerations

There are several key ethical considerations to be made when integrating advanced artificial intelligence in education. Specifically surrounding AI-enhanced education, authors have expressed concerns with the protection of sensitive student data, the potential for LLMs to embody and propagate bias, and ensuring that AI is used to enhance, and not replace, the learning process (Kasneci et al., 2023). The remainder of this section briefly describes each of these concerns and discusses them in the context of the present study.

Many of the debates surrounding data privacy with advanced AI are similar to those that have played out for educational data mining. While student data, including grades and demographic information, has been shown to contain valuable information regarding academic performance, researchers have recognized that these data are private and steps should be taken to safeguard them. (Kop et al., 2017; Yağcı, 2022). When using LLMs, for example, there are instances where students' private data could be stored for future model training. In addition, modern LLMs have memory and history features, which could pose a security risk if log-in information is shared or compromised. As such, it is considered good practice to avoid using sensitive or otherwise personal data as input to an AI systems (Wu et al., 2024). In the present work, experiments were intentionally designed to eliminate the potential for students to share private information with the LLM. Specifically, learning preferences were gleaned from student choices and profiles were generated without specific student data along the dimensions of the Felder-Silverman learning style model. As noted earlier, future work may aim to build upon the student profiles generated in this study to develop a more comprehensive personalization model; however, the use of private student data should be done with oversight and care (UNESCO, 2022).

Even though data security issues are of great importance, the ethical application of AI in education extends well beyond the pillars of data security and privacy. For instance, it is important to recognize the inherent bias present in machine learning systems that results from their training data. Like all models, LLMs are trained on a large corpus of data and, as such, their outputs are no stranger to bias (Navigli et al., 2023). This is not to say that AI cannot be integrated into K-12 education; rather, its incorporation must be carefully overseen. In the case of LLM-based PL in the classroom, this oversight is likely to come from teachers who should maintain an active role in their students' education (Kasneci et al., 2023). Given the small scale of this study and the nature of the source texts used in the rewrite model, systemic bias was not a major concern. However, as the applications of educational LLMs expand, future work should aim to recognize cases of bias and to minimize their pervasiveness in education.

Lastly, it is important to recognize that the ethical use of AI models in education requires that its end-users – namely teachers, students and, by extension of informed consent, parents/legal – be AI literate (UNESCO, 2022). Here, literacy is specifically tied to the use of AI as a learning tool and requires full transparency surrounding data sharing, data storage, and the strengths and weaknesses associated with the model's use. Such transparency is necessary as it allows users to fully understand both the accuracy and the reasoning behind why a particular output was generated. Such an understanding is important, especially when a generation does not match a user's expectations. For example, the type of data used as input enacts fundamental limits on model performance (Potgieter, 2020). Such limitations are not inherently negative, but end-users must be cognizant of them to be able to critically interpret and assess any LLM-provided output. For instance, a student should understand what the model does and does not know about them when interpreting any of its suggestions.

# 6 Conclusion

Large language models have made considerable progress in recent years (Bansal et al., 2024). These models have continued to make immense strides in text generation and, at the same time, have expanded into other media formats including the creation of images, audio and, even more recently, video (OpenAI, 2024). Owing to their sophistication, these models offer intriguing possibilities for personalized learning environments.

This randomized controlled trial was one of the first to explore this potential by using GPT-4 to glean information about and to align educational material with individual students' learning preferences. A Mann-Whitney $U$ test demonstrated that students were more likely to prefer a personalized text during the test session when it was rewritten in accordance with their profile (i.e., with their choices during the training session). This result was found to be statistically significant at the .10-level and was trending significant at the .05-level.

These findings are especially promising for the future of PL as online resources like LLMs are always available to students. Future work in LLM-powered PL environments should aim to build on the success of this study, especially given its small sample size and the limited generalizability of its results. Future work should therefore strive to involve both a larger number and wider range of students from diverse backgrounds. While this study demonstrated that GPT-4 could develop a student profile along the dimensions of the Felder-Silverman learning style model, future effort should be dedicated to expanding this profile and to designing dynamic student profiles. However, the application of AI to education must be done ethically and careful attention must be given to the design and implementation of these systems.

The present capabilities of AI presented in this study are just the beginning of their use in education. In the near-term, text-based LLMs are only expected to become more powerful while new models, like those capable of producing video, come onto the public market. As this happens, our fundamental responsibility as educators to prepare students for their future will remain unchanged. To this end, teachers must become and remain heavily involved in research to ensure all students have equal access to AI and the experience needed for their future success.

## 7 CRediT Author Contribution Statement

**MV:** Methodology, Software, Validation, Formal Analysis, Investigation, Data Curation, Writing – Original Draft, Writing – Review & Editing, Visualization. **MF:** Methodology, Software, Validation, Formal Analysis, Investigation, Data Curation, Writing – Review & Editing. **AZ:** Conceptualization, Methodology, Validation, Resources, Writing – Review & Editing, Supervision, Project Administration, Funding Acquisition.

## 8 Acknowledgements

The authors greatly appreciate the support of the school administration who made this study possible. The support of Jada Vercosa of the University of Connecticut is also appreciated.

## 9 Funding

This study was funded by a National Science Foundation (NSF) Mid-Career Advancement award, Grant No. 2120888. The first author (MV) was supported by NSF NRT No. 2152202 'TRANSCEND'.

## 10 Ethics Statement

The study procedure was reviewed and approved by the University of Connecticut's (UConn-Storrs) Institutional Review Board (IRB) under protocol number H23-0348, 31 July 2023. A parent or an otherwise legal guardian of all participants signed permission forms for their child to participate in this research study. All participants signed an assent form. The parental permission and minor assent forms used for this study were approved by the University of Connecticut's IRB.

## 11 Data Availability Statement

The de-identified, raw data supporting the conclusions of this study will be made available upon request without undue reservation. All prompts and codes used to develop and run the GPT-4 models discussed in this article are available on the first author's GitHub page at https://github.com/m-vaccaro/LLMs-and-Personalized-Learning.

## 12 Conflict of Interest

The authors declare no conflicts of interest related to this publication.